\documentclass[conference]{IEEEtran}
\usepackage[utf8]{inputenc}
\usepackage{graphicx}
\usepackage{amsmath, amssymb}
\usepackage{textgreek}
\usepackage{cite}
\usepackage{hyperref}
\usepackage{float}
\usepackage{subcaption}
\usepackage{newunicodechar}
\usepackage{tikz}
\usepackage{pgfplots}
\pgfplotsset{compat=1.17}
\usepackage{algorithm}
\usepackage{algorithmic}

\usepackage{xcolor}

\usetikzlibrary{arrows.meta, positioning}
\newunicodechar{ }{~}
\title{RL-Loop: Reinforcement Learning–Driven Real-Time 5G Slice Control for Connected and Autonomous Mobility Services}

\author{
Lara Tarkh,
Ali Chouman,
Hanan Lutfiyya,
Abdallah Shami\\[1.5mm]
\textit{The University of Western Ontario, Canada}\\
\texttt{\{ltarkh, achouman, hlutfiyy, abdallah.shami\}@uwo.ca}
}

\begin{document}
\maketitle
\begin{abstract}
Smart and connected mobility systems rely on 5G edge infrastructure to support real-time communication, control, and service differentiation. Achieving this requires adaptive resource management mechanisms that can react to rapidly changing traffic conditions. In this paper, we propose RL-Loop, a closed-loop reinforcement learning framework for real-time CPU resource control in 5G network slicing environments supporting connected mobility services. RL-Loop employs a Proximal Policy Optimization (PPO) agent that continuously observes slice-level key performance indicators and adjusts edge CPU allocations at one-second granularity on a real testbed. The framework leverages real-time observability and feedback to enable adaptive, software-defined edge intelligence. Experimental results suggest that RL-Loop can reduce average CPU allocation by over 55\% relative to the reference operating point while reaching a comparable quality-of-service degradation region. These results indicate that lightweight reinforcement learning--based feedback control can provide efficient and responsive resource management for 5G-enabled smart mobility and connected vehicle services.
\end{abstract}


\section{Introduction}

5G networks introduce the concept of network slicing, allowing multiple logical slices to share a common physical infrastructure while supporting services with diverse and often conflicting performance requirements~\cite{Chouman_5G_NWDAF, Manis_5G_Signaling}. Applications such as enhanced Mobile Broadband (eMBB), Ultra Reliable Low Latency Communications (URLLC), and massive Machine Type Communications (mMTC) operate in fundamentally different regimes. eMBB prioritizes high data rates, URLLC requires stringent latency and reliability guarantees, and mMTC focuses on scalable connectivity for large numbers of low rate devices. These heterogeneous requirements place continuous pressure on the underlying infrastructure to allocate resources in a timely, efficient, and Service Level Agreement (SLA) compliant manner.

In practice, slice performance depends on multiple resource domains. Compute capacity at user plane and edge functions determines packet processing latency, radio access network bandwidth limits the volume of traffic reaching the core, and transport and backhaul capacity directly affect end-to-end delay. A bottleneck in any of these domains can lead to noticeable performance degradation. Insufficient CPU capacity increases processing delay, limited radio resources constrain achievable throughput, and congestion in transport links increases latency~\cite{Li2020}.

Because slice demand in 5G networks can change rapidly due to user mobility and bursty application behavior, static or infrequently updated resource allocations are often inadequate. Even modest under-provisioning of compute or radio resources can result in significant latency increases for delay sensitive slices, as observed in prior work~\cite{Li2020}. These challenges motivate the need for slice controllers that operate at fine time granularity and react to real-time conditions. In this work, we focus on a single eMBB style slice with a 1~s control period. However, the proposed RL-Loop framework is designed to extend to multiple slices by applying the control policy independently using slice specific key performance indicators and normalized inputs.

A significant body of existing work on slice resource management relies on static provisioning, where allocations are chosen based on peak demand and remain fixed throughout operation. This approach simplifies deployment but wastes resources under normal load and performs poorly during sudden increases in traffic~\cite{Li2020}. Because the controller does not observe or react to short-term variations, traffic bursts often translate into QoS degradation.

Threshold-based rules are also widely used. A system might increase CPU allocation when usage exceeds a predefined percentage. Such rules are simple to configure but are reactive by design. They respond after degradation has already started and often lead to oscillatory behavior because they do not anticipate traffic dynamics~\cite{shi2020}. These mechanisms struggle in environments with irregular or rapidly changing demand.

Analytical optimization methods, including linear programming and queueing-based models, aim to compute efficient allocations but rely on assumptions such as stable load or linear traffic-response relationships~\cite{azimi2022}. These assumptions rarely hold in real deployments with mobile users and heterogeneous slices, causing decisions to become inaccurate once conditions change.

More recent predictive approaches use neural networks to forecast traffic load, such as Long Short-Term Memory (LSTM)-based autoscaling for network functions in the 5G core. While these methods can capture local patterns, they do not correct prediction errors using real-time feedback once deployed. As a result, incorrect forecasts propagate into suboptimal allocation decisions~\cite{bega2020}.

These limitations highlight the need for controllers that continuously monitor real system behavior and adapt decisions at runtime, rather than relying solely on predefined rules or offline predictions.

MicroOpt~\cite{microopt2024} represents a model-driven approach that learns a differentiable performance model from offline measurements. The model predicts QoS metrics given the current traffic and resource levels, and an optimizer uses these predictions to compute an allocation that meets the desired performance target while avoiding unnecessary resource use. This approach works well when traffic evolves slowly and the learned model remains accurate.

However, MicroOpt does not adjust its decisions based on live feedback from the system and depends entirely on the accuracy of its pre-trained model. In dynamic settings where the traffic deviates from what the model has seen during training~\cite{Manias_Model_Drift, Manias_NFV_Placement}, its allocations can become either insufficient or excessive. This limits its suitability for real-time slice control.

In this work, MicroOpt is used only offline. It provides a safe initial allocation value that a Reinforcement Learning (RL) agent starts from before deployment. Online decisions are made solely by the RL agent based on direct measurements from the 5G network. Because of this separation, the evaluation uses MicroOpt’s published QoS degradation trends only as an \emph{indicative} baseline, acknowledging that MicroOpt was originally evaluated in a simulation environment and not executed in our testbed.

Given the fast changes in slice demand and the limitations of static, threshold-based, and model-only methods, controllers must adapt their decisions directly from real-time observations. Traffic can shift quickly due to user mobility or application behavior, and different slice types impose different sensitivities to delay and throughput. A controller that depends on fixed rules or a fixed performance model may not be able to track these changes effectively.

A learning-based controller offers a more flexible alternative. By observing the actual traffic load, CPU usage, and achieved performance at each decision step, the controller can refine its decisions based on current system conditions. Reinforcement learning is particularly suited to this setting because it learns a policy that maps system states to allocation actions. Once deployed, the agent continues to act based on live measurements without requiring manual thresholds or model retraining~\cite{chen2022}. This enables QoS preservation while avoiding unnecessary resource use across a wide range of operating conditions.

We consider the problem of allocating CPU resources to 5G slice functions in order to
minimize resource usage while keeping QoS degradation $E(\beta)$ below a
target level for an eMBB-style slice. The controller observes slice-level
KPIs in real-time and updates CPU limits at a 1-s control period.

This paper makes the following contributions:
\begin{itemize}
\item  We design RL-Loop, a closed-loop controller that uses Proximal Policy Optimization (PPO) to adjust CPU allocation for a 5G slice on a 5G testbed.
\item  We show how the MicroOpt framework can be used as an offline reference,
providing a safe operating range and an \emph{indicative} baseline for QoS
degradation without re-implementing the original system.
\item We experimentally show on a 5G testbed that RL-Loop can reach a promising CPU/QoS tradeoff at a representative load, using MicroOpt only as an indicative model-driven reference point.
\end{itemize}

\section{Related Work}

Early deployments of 5G networks relied mainly on static or semi-dynamic allocation strategies. In these systems, slice resources are assigned based on peak traffic or simple configuration rules and remain unchanged for extended periods. While easy to implement, these approaches do not react to short-term variations in demand. As noted in~\cite{Li2020}, static provisioning often leads to wasted capacity under normal load and QoS degradation during traffic spikes, especially for latency-sensitive services such as URLLC and mMTC.

More recent work focuses on closed-loop control frameworks that adjust resources based on continuous monitoring. Naik et al.~\cite{closedloop2022} developed a closed-loop slice assurance mechanism that triggers automatic reallocation when performance degradation is detected. Other studies, including~\cite{bega2020, kulmar2024}, proposed predictive and hierarchical analytics approaches that use monitoring data to improve SLA adherence.

Machine learning has been widely explored for adaptive resource allocation. Deep reinforcement learning (DRL) has shown strong potential because of its ability to learn decision policies directly from environment interactions. Akyıldız et al.~\cite{Anil_Akyildiz2024-ox} showed that hierarchical DRL can allocate radio access network (RAN) resources more efficiently than static or heuristic schedulers, especially under mixed eMBB and URLLC demands. Shi et al.~\cite{shi2020} further demonstrated that DRL-based policies can dynamically optimize RAN slicing decisions under varying traffic and radio conditions. A broader view is provided in~\cite{saha2023}, which surveys RL-based slice controllers capable of managing both communication and computation resources under non-stationary traffic. Supervised learning methods have also been explored for predictive resource management and function placement in mobile networks. For example, prior work has used machine learning to anticipate user mobility and proactively improve VNF embedding decisions in B5G environments~\cite{maghe}. Other predictive approaches, including LSTM-based models, have also been proposed for autoscaling functions in the 5G core~\cite{azimi2022}. However, such prediction-driven methods do not by themselves provide a closed feedback loop and cannot correct their own decisions online once deployed. This motivates reinforcement learning approaches that continuously adapt allocations based on live system observations.

Several studies have evaluated dynamic allocation methods in domain-specific scenarios. 
Alfaqawi et al.~\cite{alfaqawi2022} analyzed 5G performance in a smart-factory environment and reported substantial improvements in throughput and latency compared to LTE, highlighting the benefits of adaptive, slice-aware control for industrial workloads. 
Xing et al.~\cite{chen2022} examined dynamic baseband and radio resource allocation for eMBB slices in heterogeneous cloud radio access network (H-CRAN) architectures and observed improvements in both throughput and QoS under varying load conditions. 
These works confirm the value of adaptive control, although many focus on a single resource type or rely on simplified test environments.

Despite recent progress, the practical adoption of AI-driven slice control still faces several challenges. Scalability and cross-domain coordination remain difficult when slices span heterogeneous RAN and core environments. Recent surveys on network slicing orchestration~\cite{rafique2025} note that current management frameworks struggle with slice lifecycle automation, performance isolation, and the integration of AI/ML components into existing network function virtualization (NFV) / software-defined networking (SDN) architectures.

Security and reliability concerns also persist. As networks move toward zero-touch automation, AI-based controllers become part of the attack surface. Batewela et al.~\cite{sadeep2025} highlight that orchestration functions introduce new vulnerabilities related to real-time monitoring, intent translation, and policy enforcement. Ensuring stable and explainable behavior from learning-based controllers remains an open problem, particularly under non-stationary traffic or adversarial conditions.

Existing work explores both model-driven and RL-based methods, but most solutions treat these directions separately and rarely evaluate them in a full end-to-end 5G testbed. In this paper, we introduce RL-Loop, a closed-loop controller that uses real-time feedback to adjust CPU allocation for slice functions. MicroOpt is used only offline to provide an initial reference point, while all online decisions are made by the RL agent operating directly on network measurements. This combination addresses limitations in prior static, reactive, and prediction-only methods and provides a practical framework for adaptive slice control.
\section{Methodology}
\label{methodology}

RL-Loop runs as a closed-loop between a reinforcement learning agent and a 5G testbed. The agent observes slice-level metrics from the testbed and adjusts CPU allocation in real-time. The goal is to keep QoS behavior stable while avoiding unnecessary over-provisioning.

At each control step, the agent receives a vector with traffic and CPU information from the testbed. It then outputs a new CPU allocation value that is applied to the slice. The next observation reflects the effect of this action, and the agent updates its policy over time based on the rewards it receives.

\begin{figure}[t]
\centering

\begin{tikzpicture}[
    node distance=1.4cm and 1.8cm,
    every node/.style={font=\small},
    box/.style={
        draw,
        rounded corners,
        minimum width=3.0cm,
        minimum height=1.0cm,
        align=center,
        thick
    },
    arrow/.style={-{Latex[length=2mm]}, thick}
]

\definecolor{metricColor}{RGB}{255,236,204}
\definecolor{testbedColor}{RGB}{220,230,255}
\definecolor{agentColor}{RGB}{210,245,210}

\node[box, fill=metricColor] (metrics) {Metrics / KPIs\\(traffic, CPU, throughput)};
\node[box, fill=testbedColor, below=1.6cm of metrics] (testbed) {Testbed Environment\\(Open5GS + UERANSIM)};
\node[box, fill=agentColor, right=2.2cm of metrics] (agent) {RL Agent\\(PPO policy)};

\node[
    draw,
    dashed,
    rounded corners,
    above=1.6cm of agent,
    minimum width=3.3cm
] (microopt) {MicroOpt offline reference};

\draw[arrow] (metrics) -- node[above]{State} (agent);
\draw[arrow] (agent) -- node[midway, right]{CPU limit action} (testbed);
\draw[arrow] (testbed) -- node[midway,left]{Observation} (metrics); 

\draw[densely dotted, -{Latex[length=2mm]}]
    (microopt) -- node[right]{initial CPU range} (agent);

\end{tikzpicture}
\caption{RL-Loop closed feedback loop in the 5G testbed. The RL agent receives KPIs as state and sends CPU limit update actions directly to the testbed every second.}
\label{fig:rlloop-architecture}
\end{figure}
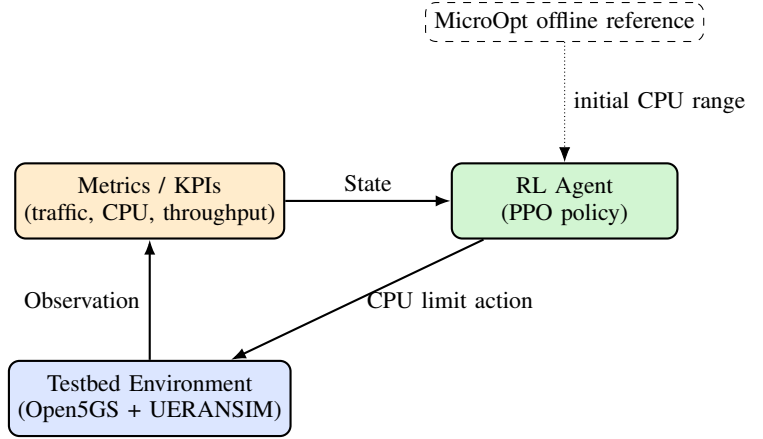

MicroOpt is not part of the online loop in our implementation. It is used only as a reference point. Its published operating region and QoS trends guide the choice of a safe CPU range and serve as an indicative baseline in the evaluation, but the RL agent itself interacts directly with the testbed without calling MicroOpt during runtime.
The overall feedback architecture is shown in Fig.~\ref{fig:rlloop-architecture}.

\begin{figure*}
    \centering
    \includegraphics[width=\linewidth]{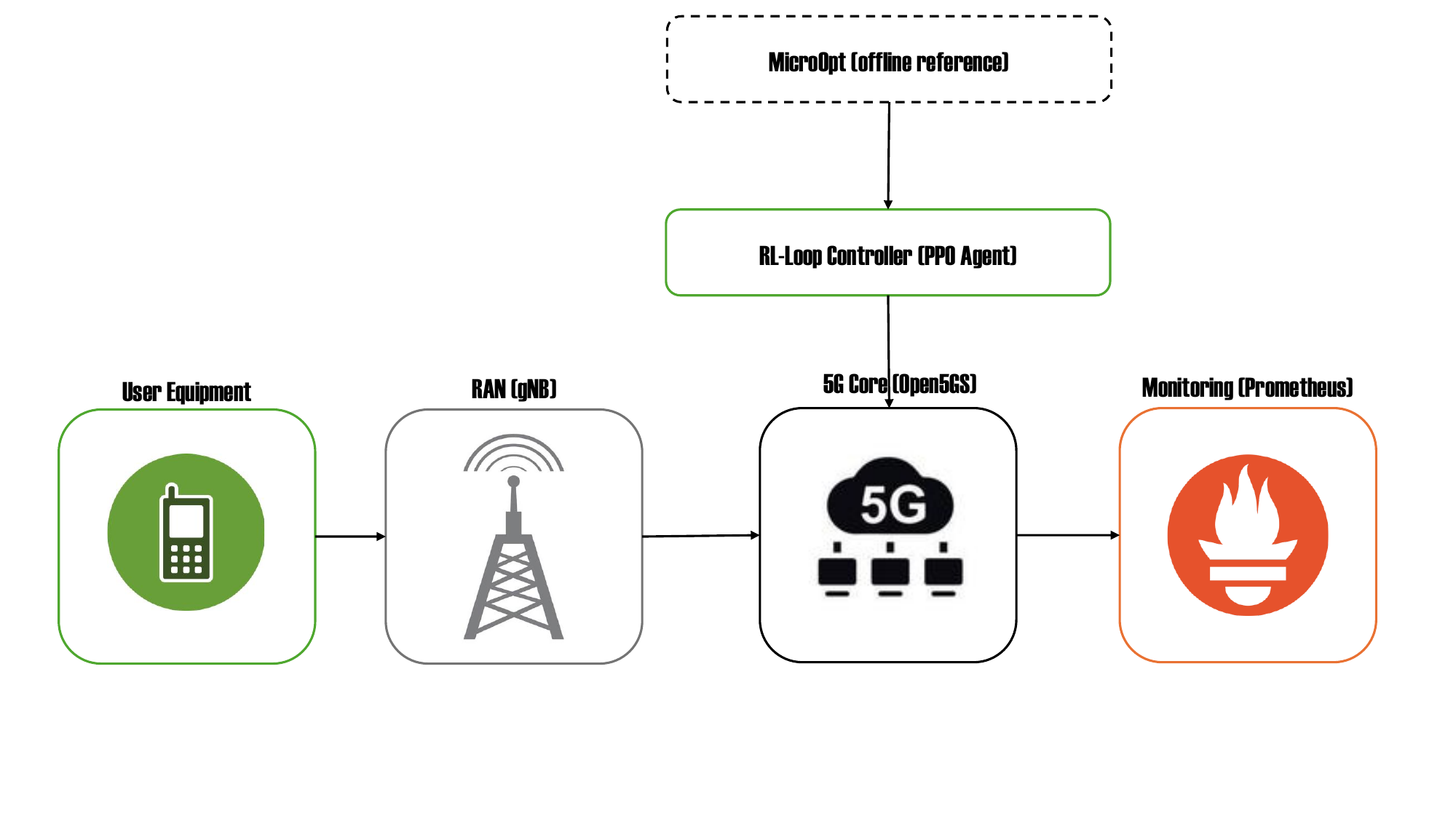}
    \caption{5G testbed used for RL-Loop experiments}
    \label{testbedfig}
\end{figure*}

All experiments are run on an end-to-end 5G testbed built with Open5GS for the core network and UERANSIM for the radio access network and user equipment, as shown in Fig.~\ref{testbedfig}. The core network functions, gNB, and UE instances run as virtual machines on a Hyper-V host with an Intel Xeon Gold 6348 processor and 512 GB of RAM. Each network function is deployed on Ubuntu 24.04 in a private network.

Traffic is generated by emulated UEs that stream a 4K video over the 5G slice. User arrivals follow a truncated normal distribution with an average between one and five users per second, with parameters chosen to emulate realistic workload fluctuations inspired by the Telecom Italia Big Data Challenge dataset for Milan and Trentino~\cite{italia}. This creates realistic variations in offered load while remaining controllable.

For each run, we select a CPU limit and a bandwidth setting for the slice and keep them fixed while the traffic evolves.
At each one-second interval, the system logs four values: the CPU usage of the user-plane function (in millicores), the throughput observed at the UE side (computed from packet captures using \texttt{tshark}), the CPU allocation chosen by the RL agent, and the corresponding reward value returned by the environment.

These logs are stored as CSV and text files and are used both to train the agent offline and to compute aggregate metrics such as average CPU allocation and qualitative QoS trends.

MicroOpt is a model-driven optimization framework that was introduced in prior work~\cite{microopt2024}. It learns a differentiable performance model that maps traffic load and resource levels to a predicted QoS value. An optimizer then uses this model to select resource allocations that satisfy a QoS target while keeping resource usage low. In its original evaluation, MicroOpt uses traffic profiles derived from the Telecom Italia dataset and runs in a simulation environment.

In this paper, we do not reimplement or execute MicroOpt on our testbed. Instead, we use the published MicroOpt results as an indicative baseline. In particular, we rely on the reported QoS degradation curves and CPU operating points under similar traffic ranges. These numbers provide a useful reference to understand where our RL policy sits relative to a model-driven controller, while we clearly state that the comparison is not one to one because the environments differ.
Methodologically, our use of MicroOpt as an indicative baseline is similar in spirit to the comparison strategy in~\cite{indicative}, where a real testbed controller is evaluated against baseline results obtained in a separate simulation environment. In both cases, the comparison is framed explicitly as non one-to-one, but still useful to position the proposed controller relative to prior work.

At initialization, the RL agent starts from a CPU value of 0.5 cores, which translates to 500~millicores from the container-based CPU allocation. The action space for CPU allocation, which is the safe operating region suggested by MicroOpt, is in the range of 0.5 - 4 cores at increments of 0.5 cores each.
\subsection{Reinforcement Learning Framework}

The RL controller is trained with the Proximal Policy Optimization (PPO) algorithm using Stable Baselines3, an open-source Python library for reinforcement learning. The goal of the agent is to align CPU allocation with an online estimate of slice load using measurements that are observable in the testbed.
For each slice, the observation vector contains two normalized entries: the current traffic load and the CPU usage, both scaled to the interval $[0,1]$ using fixed min–max ranges from the testbed configuration.

The action space is a single continuous value in \([0,1]\) that represents the normalized CPU allocation for the slice. This value is mapped back to a concrete CPU limit in millicores when applied on the testbed. A higher action value gives more CPU to the slice, while a lower value reduces its share.

 The reward function encourages the agent to keep allocation close to an online estimate of slice load. At time step \(t\), the reward is  \begin{equation} r_t = 1 - \frac{1}{n} \sum_{i=1}^{n} \left| a_i - d_i \right|, \end{equation}  where \(n\) is the number of slices, \(a_i\) is the normalized CPU allocation for slice \(i\), and \(d_i\) denotes a normalized load estimate for the slice, computed online from the current number of active UEs and their target throughput requirements, then scaled to the interval \([0,1]\) using fixed min--max bounds. In our testbed, we use a single slice, so \(n = 1\). The reward is equal to one when allocation matches this normalized load estimate exactly and decreases as the gap between them grows. This penalizes both over-allocation and under-allocation.

Training starts offline, using measurement traces collected from the testbed under different CPU settings and loads. This allows the agent to see a wide range of state action pairs before it is deployed. We then switch to online training, where the agent interacts directly with the live testbed and continues to refine its policy based on real feedback.

\subsection{RL-Loop Online Control Algorithm}

Algorithm~\ref{alg:rlloop} outlines the interaction between the PPO agent and 
the 5G testbed. The controller operates with a 1~s control period: at each step, 
the environment reports the current traffic load, CPU usage, and throughput. The 
agent generates a normalized CPU allocation and the environment applies the 
corresponding CPU limit to the user-plane function. The next observation reflects 
the effect of this action, and the reward is computed based on the allocation--load-estimate gap. This closed-loop interaction continues throughout the run and provides the 
experience used for the PPO policy updates.

\begin{algorithm}[t]
\caption{RL-Loop Online CPU Control (PPO-based)}
\label{alg:rlloop}
\begin{algorithmic}[1]

  \STATE \textbf{Inputs }Environment $\textit{Env}$, PPO policy $\pi_\theta$, value function $V_\phi$, control period $\Delta t = 1$~s
  \STATE\textbf{Output } Trained policy $\pi_\theta$

  \STATE Initialize $\theta$, $\phi$, and state $s_0 \gets \textit{Env.reset}()$

  \FOR{episode $=1$ to $N_{\text{episodes}}$}
    \STATE Clear experience buffer $B$
    \FOR{$t = 0$ to $T-1$}
      \STATE Observe KPIs from env \& build normalized state $s_t$
      \STATE Sample action $a_t \sim \pi_\theta(\cdot \mid s_t)$
      \STATE Apply CPU limit corresponding to $a_t$ through $\textit{Env.applyCpuLimit}(a_t)$
      \STATE Wait $\Delta t$ and obtain next state $s_{t+1}$
      \STATE Compute reward $r_t$
      \STATE Store transition $(s_t, a_t, r_t, s_{t+1})$ in $B$
    \ENDFOR

    \STATE Compute advantages and returns from $B$
    \STATE Update $\theta$ and $\phi$ using PPO on minibatches from $B$
  \ENDFOR

\end{algorithmic}
\end{algorithm}

\subsection{Feedback Loop Design}

During online operation, the RL agent and the testbed form a closed-loop. The loop runs with a control period of one second.

At each step, the environment collects the current traffic load, CPU usage, and throughput from the testbed to build the observation vector. The agent then receives this observation and outputs a new normalized CPU allocation. The environment converts this action into a CPU limit in millicores and applies it to the user-plane process through Linux cgroup controls, which enforce per-process CPU resource limits at runtime. In the next step, the environment measures the updated CPU usage and computes the reward based on the difference between allocation and the normalized load estimate.

This cycle repeats for the duration of the experiment. Because the agent sees the effect of its own actions in the next observation, it can gradually correct mistakes and learn how quickly the system reacts to allocation changes.

QoS constraints are not encoded as hard rules in the environment. Instead, they are reflected indirectly in the reward design and in the chosen load-estimate ranges. When the agent allocates too little CPU, throughput drops while the normalized load estimate remains high, which lowers the reward. When it allocates too much, CPU usage grows without a corresponding benefit and the reward also decreases. The agent therefore learns to stay in a region where CPU headroom and slice performance are balanced.

\subsection{Implementation Details}

The RL environment and agent are implemented in Python. We use Stable Baselines3 with the PPO algorithm and the default multilayer perceptron policy. The code runs on a Linux controller node that has network access to the 5G core and gNB virtual machines.

The main PPO hyperparameters are a learning rate of \(3 \times 10^{-4}\), a discount factor \(\gamma = 0.99\), and a batch of 32 time steps per update. The agent interacts with the testbed for 900 steps, which corresponds to 15 minutes of wall clock time at one second per action.

Inference latency is small compared to the control period. Each forward pass of the policy network takes less than ten milliseconds on a standard CPU. This is negligible relative to the one second interval between control decisions and fits within near real-time constraints.

In the evaluation, we report the average CPU allocation achieved by RL-Loop on the testbed and compare it to the QoS degradation trends and CPU points published for MicroOpt. The MicroOpt numbers are used only as an indicative baseline.

\section{Evaluation}

\subsection{Experiment Setup}

We evaluate RL-Loop on the end-to-end 5G testbed described in Section~\ref{methodology},   built with Open5GS for the core network and UERANSIM for the radio access network (RAN) and user equipment (UE) emulation~\cite{10559989}. All core functions, the gNB, and UE instances run as virtual machines on a Hyper-V host with an Intel Xeon Gold 6348 CPU and 512~GB of RAM. Each VM runs Ubuntu~24.04~LTS in a private network.

Traffic is generated by emulated UEs streaming a 4K YouTube video~\cite{Tejulaify_undated-om}. User arrivals follow a truncated normal distribution with a mean between 1 and 5~users/second, using statistics inspired by the Telecom Italia trace~\cite{italia}. This produces realistic but controllable variations in offered load.

During each run, the RL agent interacts with the testbed for 900~seconds. At every one second control interval it receives the current traffic and CPU usage, selects a normalized CPU allocation, and the environment applies the corresponding CPU limit (in millicores) to the user plane process using Linux cgroup controls.
We record four quantities at one-second resolution: CPU usage of the user-plane function (in millicores), UE-side throughput extracted from packet captures using \texttt{tshark}, the CPU allocation selected by the RL agent, and the reward returned by the environment.

All logs are stored as CSV or text files and are used to compute the mean CPU allocation over the run and to study how throughput and reward evolve over time.

\subsection{Performance Metrics}

We focus on two main metrics that directly reflect the goals of RL-Loop.

\subsubsection{Average CPU Allocation}
This is the mean CPU limit applied to the user plane function during the run, expressed in millicores. It captures how much compute capacity is reserved by the controller.

\subsubsection{QoS Degradation}
To align with the MicroOpt framework~\cite{microopt2024}, we adopt the same
definition of QoS degradation used in their slice optimization model. For a
throughput trace $q(t)$ and a slice-level throughput threshold $q_{\text{thresh}}$,
the degradation over a reconfiguration interval is defined as the
traffic-weighted proportion of time during which the QoS falls below the
threshold:
\[
\beta = 
\frac{\sum_{t=1}^{\tau} x(t)\,1\!\left[q(t)\le q_{\text{thresh}}\right]}
     {\sum_{t=1}^{\tau} x(t)},
\]
where $x(t)$ is the slice traffic (users/second), $1[\cdot]$ is the indicator
function, and $\tau$ is the horizon length. This quantity is used in
MicroOpt's primal–dual optimization as a constraint on the expected QoS
degradation. Although MicroOpt replaces the indicator by a smooth surrogate
during gradient-based updates, the strict expression above corresponds to the
actual degradation metric evaluated in their study.

In our testbed evaluation, we follow the same definition by computing
$1[q(t)\le q_{\text{thresh}}]$ over all one-second samples of measured UE
throughput, using $q_{\text{thresh}}=1$~Mbps for the eMBB slice. The resulting
$\beta$ therefore provides a degradation metric that is consistent with the
MicroOpt formulation, allowing us to use MicroOpt's published degradation
curves as an indicative baseline, even though their evaluation is performed in
simulation while RL-Loop operates on a real testbed.

In our testbed, we also monitor a simple SLA view by counting the fraction of samples where throughput stays above 1~Mbps, which we use as a target for eMBB slice use case. This supports the interpretation of $E(\beta)$ but the main comparison to MicroOpt uses the same degradation metric.

For context, we refer to the CPU operating points and QoS degradation curves reported for MicroOpt in its original simulation study~\cite{microopt2024}. These values are used only as an \emph{indicative} baseline, since MicroOpt is not executed on our testbed.

\subsection{Initial Evaluation}

\begin{figure}[t]
    \centering
    \includegraphics[width=\linewidth]{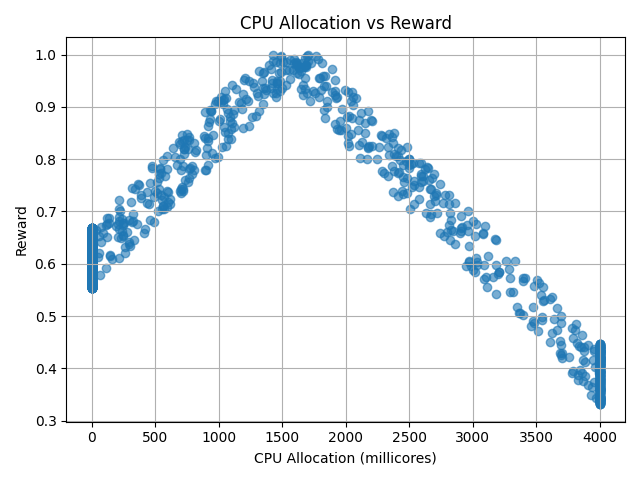}
    \caption{Reward versus CPU allocation for RL-Loop. The reward is highest at intermediate CPU levels, where allocation best matches the normalized load estimate.}
    \label{fig:cpualloca_reward}
\end{figure}

We first focus on a 5~users/second arrival rate, which matches the operating point reported for MicroOpt in~\cite{microopt2024}.
Figure~\ref{fig:cpualloca_reward} shows the relation between the CPU allocation selected by the agent and the resulting reward. The reward peaks around a normalized allocation of $0.3$--$0.4$, which confirms that the reward function penalizes both over-allocation and under-allocation and encourages the agent to track the normalized load estimate.
Under the 5~users/second arrival rate, RL-Loop converges to a stable operating point in our testbed runs.

Over the 15-minute run, the mean normalized CPU allocation selected by RL-Loop is approximately $0.33$, which corresponds to an average of \textit{1330.76~millicores}. For comparison, the MicroOpt study reports a CPU level of \textit{2945.72~millicores} for a 5~users/second load in the same traffic range~\cite{microopt2024}. Using these values as context, RL-Loop operates at about \textit{45\%} of the CPU used by MicroOpt, i.e., a reduction of slightly more than \textit{55\%} in average CPU allocation. Using the same definition of $E(\beta)$ as MicroOpt and applying it to our testbed traces, we obtain a QoS degradation value of \textit{0.10} for RL-Loop at 5~users/second. This matches the degradation reported for MicroOpt under the same arrival rate in its simulation environment~\cite{microopt2024}. Table~\ref{tab:cpu_qos_comparison} summarizes the corresponding CPU values and degradation metrics.

\begin{table}[t]
\centering
\caption{CPU usage and QoS degradation at 5~users/second. 
MicroOpt values are taken from~\cite{microopt2024} and used as an indicative baseline.}
\label{tab:cpu_qos_comparison}
\begin{tabular}{|l|c|c|}
\hline
\textbf{Approach} & \textbf{CPU (millicores)} & \textbf{$E(\beta)$} \\
\hline
MicroOpt (simulation) & 2945.72 & 0.10 \\
RL-Loop (testbed)     & 1330.76 & 0.10 \\
\hline
\end{tabular}
\end{table}
While practical controller baselines such as static allocation, threshold-based autoscaling, and proportional control are important for a full operational comparison, implementing and tuning these baselines on the physical Open5GS+UERANSIM testbed is left to future work. In the current version, MicroOpt is used only as an indicative reference point rather than as a fully matched practical control baseline. Accordingly, our evaluation should be interpreted as demonstrating the feasibility and promise of RL-Loop on a real testbed, rather than as a definitive comparison against all conventional controllers. 
In our logs, throughput remains above 1~Mbps for the majority of samples, with short drops that correlate with sharp changes in traffic. Because of limitations in packet capture granularity and virtualized I/O, we treat these SLA observations as supportive but not as a stricter metric than $E(\beta)$.

Overall, these results suggest that RL-Loop can maintain a QoS degradation level comparable to the MicroOpt operating point while using substantially less CPU. The comparison remains indicative, since MicroOpt was evaluated in simulation and RL-Loop runs on a real testbed, but it shows that learning-based closed-loop control can reach the same qualitative SLA region with much lower compute usage.
\subsection{RL Training Behaviour}

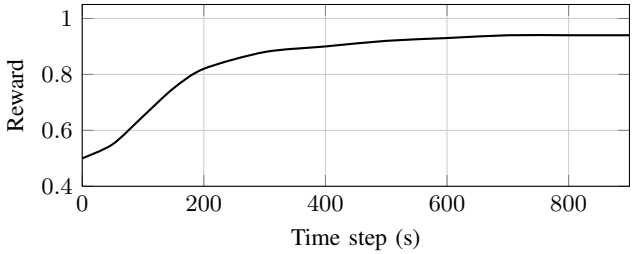
\begin{figure}[t]
    \centering
    \begin{tikzpicture}
    \begin{axis}[
        width=\linewidth,
        height=4cm,
        xlabel={Time step (s)},
        ylabel={Reward},
        xmin=0, xmax=900,
        ymin=0.4, ymax=1.05,
        xtick={0,200,400,600,800},
        ytick={0.4,0.6,0.8,1.0},
        grid=both,
        grid style={line width=0.3pt, draw=gray!40},
        every axis/.append style={font=\small},
    ]
        \addplot[
            smooth,
            thick
        ] coordinates {
            (0,   0.50)
            (50,  0.55)
            (100, 0.65)
            (150, 0.75)
            (200, 0.82)
            (300, 0.88)
            (400, 0.90)
            (500, 0.92)
            (600, 0.93)
            (700, 0.94)
            (800, 0.94)
            (900, 0.94)
        };
    \end{axis}
    \end{tikzpicture}
    \caption{Evolution of the RL-Loop reward during the 900~s online run.}
    \label{fig:reward_time}
\end{figure}

Figure~\ref{fig:reward_time} shows the evolution of the reward during the
900~s online run. The instantaneous reward is noisy at the beginning,
when the PPO agent is still exploring different CPU levels. After the
first 200--300~steps, the moving average converges to a stable range
around 0.8–1.0 and remains there for the rest of the experiment. This
indicates that the policy has learned a consistent mapping between
traffic/CPU measurements and allocation decisions and does not exhibit
large oscillations.

Combined with Fig.~\ref{fig:cpualloca_reward}, which shows that the
highest rewards are obtained around 1200–1600~millicores, this suggests
that the agent has discovered an efficient operating region: it avoids
both persistent under-allocation (which would sharply reduce the
reward) and persistent over-allocation (which offers no additional
benefit but still incurs a penalty). In other words, the controller
converges to a stable policy that tracks the normalized load estimate while keeping CPU usage low.

\section{Discussion}

\subsection{Performance Analysis}

The RL agent is able to operate stably in a closed-loop with the 5G testbed. Figure~\ref{fig:cpualloca_reward} shows the reward as a function of the CPU allocation selected by RL-Loop. At low CPU levels, the reward remains small, reflecting under-provisioning when the normalized load estimate exceeds the allocated compute. As the allocation increases, the reward rises and peaks around 1200--1600~millicores, corresponding to the region where the CPU best matches the normalized load estimate. Beyond this range, the reward decreases again as additional CPU does not provide a corresponding benefit and is penalized as waste. The average allocation chosen by the learned policy (1330.76~millicores) lies inside this high-reward band, which confirms that the agent has learned to avoid both under-allocation and over-allocation and to operate near the efficient operating point. PPO is suited for this setting because it provides stable policy updates in
continuous control problems, where overly aggressive updates can destabilize the
feedback loop. In the methodology, PPO’s clipped objective ensures that the
policy evolves smoothly as the agent observes new KPIs, preventing sudden jumps
in CPU allocation that could degrade slice performance. This behaviour is
consistent with what we observe in Fig.~\ref{fig:cpualloca_reward} and
Fig.~\ref{fig:reward_time}, where the agent converges to a narrow operating
region and avoids oscillations. In contrast, value-based agents such as DQN are
not applicable to continuous action spaces without discretization, and
policy-gradient methods without clipping (e.g., vanilla REINFORCE) tend to
produce high-variance updates that lead to unstable allocations in online
control. The stable convergence of the PPO policy in our testbed therefore
matches its expected advantages and supports its use for real-time slice
resource control.

Using the MicroOpt degradation metric, RL-Loop achieves $E(\beta)=0.10$, which matches the value reported for MicroOpt at 5~users/second in its simulation study~\cite{microopt2024}. At the same time, RL-Loop uses on average about 45\% of the CPU that MicroOpt reserves in that operating region. This supports the claim that a learning-based controller driven by real testbed feedback can preserve SLA trends while substantially lowering resource usage.

The absolute throughput values observed on the testbed are influenced by infrastructure constraints such as virtualized network I/O and the way packet captures are collected. For this reason, we focus on relative behaviour: throughput stays above a simple 1~Mbps SLA threshold for most of the run, and the resulting degradation metric aligns with the MicroOpt reference, despite the different environments.

\subsection{Limitations and Challenges}
The current prototype also reveals several limitations. The testbed environment introduces inherent delays between applying a new CPU limit and observing its full effect, which can blur the link between action and reward. In addition, the traffic model relies on a single video-streaming workload and a single slice, so future multi-slice scenarios may exhibit different behaviors. Finally, throughput and SLA statistics are derived from coarse packet captures, which omit finer QoS aspects such as jitter or application-level quality.
These limitations affect the precision of our QoS analysis but do not change the main observation that RL-Loop is able to operate at a much lower CPU level than the MicroOpt operating point while remaining in a similar degradation regime.

\subsection{Future Work}

A natural next step is to port MicroOpt to the same testbed and obtain a fully aligned baseline. This would isolate the effect of the learning-based controller from environment differences and provide a stricter quantitative comparison, but it is orthogonal to the main contribution of this paper, which is to demonstrate that an RL-based feedback loop can operate stably on a real 5G testbed and significantly reduce CPU usage. Future work will focus on improving observability with finer-grained traffic and CPU monitoring to obtain cleaner QoS measurements, extending the action space to jointly control CPU, bandwidth, and possibly buffer sizes or scheduling priorities, and exploring multi-slice and multi-agent extensions where several RL controllers coordinate allocations across slice types. Future work will also include comparisons against practical baselines such as static allocation, threshold-based autoscaling, and proportional control, as well as reward formulations based only on directly measurable KPIs.

\section{Conclusion}

We presented RL-Loop, a reinforcement learning-based feedback controller for 5G slice resource allocation. RL-Loop observes live KPIs from an Open5GS+UERANSIM testbed and adjusts CPU limits every second using a PPO policy.

In contrast to static, threshold-based, or purely model-driven approaches, RL-Loop learns its behaviour directly from measurements and continues to adapt after deployment. Using MicroOpt’s published results as an indicative baseline, our initial evaluation shows that RL-Loop can operate with more than 50\% lower average CPU allocation while maintaining acceptable qualitative QoS behaviour on the testbed.

These results suggest that combining simple closed-loop RL with existing model-driven insights is a practical way to improve resource efficiency in 5G slicing. A more extensive evaluation, including running MicroOpt on the same testbed and handling multiple slices, is left for future work.
\bibliographystyle{IEEEtran} \bibliography{references}
\end{document}